\documentstyle[aps,pra,epsfig,twocolumn,graphics]{revtex} 

\def\be{\begin{equation}}
\def\ee{\end{equation}}
\def\bea{\begin{eqnarray}}
\def\eea{\end{eqnarray}}
\def\bma{\begin{mathletters}}
\def\ema{\end{mathletters}}
\def\tr{{\rm tr}}

\def\C{\hbox{$\mit I$\kern-.7em$\mit C$}}

\newcommand{\one}{\mbox{$1 \hspace{-1.0mm}  {\bf l}$}}
\newcommand{\ket}[1]{ | \, #1  \rangle}
\newcommand{\bra}[1]{ \langle #1 \,  |}
\newcommand{\proj}[1]{\ket{#1}\bra{#1}}

\begin{document}
\draft

\title{Optimal Creation of Entanglement Using a Two--Qubit Gate}

\author{B. Kraus and J.I. Cirac}

\address{ Institute for
Theoretical Physics, University of Innsbruck,
A--6020 Innsbruck, Austria}

\date{\today}

\maketitle
\begin{abstract}
We consider a general unitary operator acting on two qubits in a
product state. We find the conditions such that the state of the
qubits after the action is as entangled as possible. We also
consider the possibility of using ancilla qubits to increase the
entanglement.
\end{abstract}

\narrowtext

\section{Introduction}

Entanglement \cite{EPR,Sch} is a quantum
mechanical feature that can be employed for
computational and communication purposes. During
the last few years a big effort has been done in
order to create entanglement in several
laboratories \cite{fo00}. This entanglement can
then be used for many fascinating things such as
teleportation \cite{Bennett_tel}, quantum
cryptography \cite{Ekert} and quantum computation
\cite{comput}. In some of these experiments the
entanglement is produced by starting out from a
product state of two systems (typically qubits)
and using some physical process that gives rise
to an interaction between them. Thus, one of the
relevant problems in this context is to find ways
of generating "as much entanglement as possible"
for a given experimental set--up, i.e. a
non--local interaction.

The first steps to answer this problem have been given in Ref.
\cite{Za00,Du00,Ci00}. In particular, given a non--local
Hamiltonian D\"{u}r {\it et al} have found the optimal way of
generating entanglement. It consists of applying some fast local
operations {\it during} the interaction processes in such a way
that the rate at which entanglement increases is always maximal.
In some situations, however, one cannot apply fast local
operations during the process, but rather a fixed quantum gate
is given. In this work we find the states $\ket{\phi}_A$ and
$\ket{\psi}_B$ for which the entanglement of
$U_{AB}\ket{\phi}_A\ket{\psi}_B$ is maximal, where $U_{AB}$ is
an arbitrary unitary operator. Thus, our results give a
characterization of two--qubit gates in terms of the
entanglement that they can produce. For example, we will
determine which are the operators $U_{AB}$ that can create
maximally entangled states. While most of our results are
concerned with two qubits, we will also show that if we allow
them to be initially (locally) entangled with some ancillas, one
can obtain more entanglement, at least, for certain measures of
entanglement.

In general, an arbitrary unitary operator acting on two qubits
can be parameterized in terms of 15 coefficients (plus a global
phase). Thus, to study the maximum entanglement which can be
produced in terms of all these parameters seems a formidable
task. However, we will show that one can always decompose
$U_{AB}= (U_A\otimes U_B)U_d(V_A\otimes V_B)$, where $U_d$ has a
special form that only depends on three parameters and the rest
are local unitary operators. This implies that we can restrict
ourselves to characterize operators in the form $U_d$. The use
of the magic basis introduced in Ref. \cite{Wo97} will also
considerably simplify our derivations.

This paper is divided into four sections. In Section II we
introduce our notation and recall some measures of entanglement
and their properties. In Section III we show that there exists a
decomposition of any two--qubit gate, which allows us to
simplify the problem. In Section IV we consider the problem of
two qubits. We determine how much entanglement can be produced
by a general two--qubit gate acting on a product state. We also
find which of such states give rise to that amount of
entanglement. In Section V we discuss the case where we allow
the qubits to be initially entangled with ancillas. We will show
that the solution to the problem depends on the measure of
entanglement we use to quantify it. We will also give two
examples in which this problem can be solved analytically for a
particular measure of entanglement.

\section{Definitions and properties of entanglement measures}

The purpose of this Section is two fold. On the one hand, we
give the definitions and notations that will be used throughout
the whole paper. On the other hand, we review some measures of
entanglement (for pure states) and some of their properties.

\subsection{Definitions}

We consider two partners, Alice and Bob, who posses two quantum
systems, $A$ and $B$, respectively. These systems will be
composed of one or two qubits each. We will express the states
of these qubits in terms of the computational basis,
$\{\ket{0},\ket{1}\}$. The Hilbert space of system $A$ ($B$)
will be denoted by ${\cal H}_A$ (${\cal H}_B$) respectively.

Throughout this paper we use capital Greek letters for joint
states of systems $A$ and $B$ and small letters for states
describing either system $A$ or system $B$. We denote by
$\ket{\Psi^\perp}$ a state which is orthogonal to $\ket{\Psi}$,
whereas $\ket{\Psi^\ast}$ denotes the complex conjugate of
$\ket{\Psi}$ in the computational basis. We will denote the
Pauli operators by $\sigma_x, \sigma_y, \sigma_z$ and by
$\vec{\sigma}=(\sigma_x,\sigma_y,\sigma_z)$. If it is not clear
which system an operator is acting on, we specify it with either
a sub-- or superscript, e.g. $\vec{\sigma}_A$ or $\sigma_x^A$.

For two qubits, the Bell basis is defined as
follows: 
\bea
\ket{\Phi^\pm}=\frac{1}{\sqrt{2}}(\ket{00}\pm\ket{11}),
\ket{\Psi^\pm}=\frac{1}{\sqrt{2}}(\ket{01}\pm\ket{10}).
\eea 
We also make use of the so--called magic
basis \cite{Wo97}, which is defined in the same
way as the Bell-basis except for some global
phases. We will denote the elements of this basis
by 
\bma 
\bea 
\ket{\Phi_1} &=& \ket{\Phi^+}, \quad
\ket{\Phi_2}=-i\ket{\Phi^-},\\
\ket{\Phi_3}&=&\ket{\Psi^-}, \quad
\ket{\Phi_4}=-i\ket{\Psi^+}. 
\eea 
\ema 
The
coefficients of a general state in that basis
will be typically denoted by $\mu_k$; that is, we
write $\ket{\Psi}= \sum_k \mu_k \ket{\Phi_k}$. In
what follows $\ket{\Phi}$ denotes a maximally
entangled state.

\subsection{Measures of entanglement}

We review here some measures of entanglement for
pure states. In the first part of the paper we
are going to use the so--called {\it concurrence}
\cite{Wo972}, $C$. It is defined as
\bea
C(\ket{\Psi})=|\bra{\Psi}\sigma_y\otimes \sigma_y
|\Psi^\ast\rangle|.
\eea
Writing $\ket{\Psi}$ in
the magic basis we get
\bea
C(\ket{\Psi})=|\sum_k\mu_k^2|.
\eea

In the second part of the paper we will use other
measures of entanglement, which are better
expressed in terms of the Schmidt coefficients. A
pure state $\ket{\Psi}$, describing the state of
two particles, $A$ and $B$, each of dimension $m$
always has a Schmidt decomposition in the form:
\bea 
\ket{\Psi}=\sum_{k=1}^{m} c_k
\ket{\phi_k}_A\ket{\psi_k}_B ,
\eea
where $\bra {\phi_k}\phi_l\rangle=\bra
{\psi_k}\psi_l\rangle=\delta_{kl}$ $\forall
k,l=1,\ldots ,m$. The real and positive
coefficients $c_k$, which are the square roots of
the eigenvalues of the reduced density operator,
$\rho_A=\tr_B (\proj{\Psi})$ (or
$\rho_B=\tr_A(\proj{\Psi})$) are called Schmidt
coefficients. We will choose them in increasing
order, i.e. $c_1\leq c_2 \ldots\leq c_m$.

The {\it Entropy of entanglement} is defined as
follows: 
\bea 
E_E(\ket{\Psi})\equiv
S(\rho_A)=-\tr[\rho_A\log_2(\rho_A)]=-\sum_{k=1}^m
c_k^2 \log_2 (c_k^2). 
\eea 
This measure has a
well defined meaning: given $n$ copies of a state
$\ket{\Psi}$ then one can produce, using only
local operations and classical communication,
$nE_E(\ket{\Psi})$ maximally entangled states and
vice versa (in the limit $n\rightarrow \infty$).

Another useful measure is the {\it Schmidt number} \cite{Peres},
which we will denote by $E_S$. It is the number of Schmidt
coefficients which are different than zero minus one.

There are other set of measures, the so--called
{\it Entanglement Monotones}, which arise in the
context of allowed modification of entangled
states under local operations \cite{Vi98}. They
are defined as
\bea
E_n(\ket{\Psi})=\sum_{k=1}^{n} c_k^2,
\eea for
$n=1,...,m-1$.

We will also use the so--called 2--Entropy (or
2--R\`{e}nyi entropy) of the reduced density
operator \cite{Ho96}. It is defined as 
\bea
E_R(\ket{\Psi})\equiv
S_R(\rho_A)=1-\tr(\rho_A^2)=1-\sum_{k=1}^m c_k^4,
\eea 
where again $\rho_A$ denotes the reduced
density operator of $\proj{\Psi}$ and $c_k$ the
Schmidt coefficients of $\ket{\Psi}$. In the
following we will call this measure the {\it
R\`{e}nyi entanglement}.

\subsection{Properties}

A state describing two qubits contains two Schmidt coefficients
at most, $c_1,c_2$, where $c_2=\sqrt{1-c_1^2}$. Thus, its
entanglement is completely determined by one parameter, $c_1$.
All the measures of entanglement are monotonic functions of each
other and, therefore, equivalent. In higher dimensions
($m$-level systems) though, this is no longer true. Let us
denote now by $\ket{\Psi}$ and $\ket{\Psi'}$ two states
describing two $m$-level systems. Then it might happen that for
some measure $\ket{\Psi}$ is more entangled than $\ket{\Psi'}$,
whereas for some other measure it is the other way around.

Let us briefly recall some of the properties which have to be
satisfied by any measure of entanglement, $E$ \cite{Ho99}:

\begin{description}
\item [(a)] Monotonicity under local operations: Suppose that Alice
makes a measurement on her qubit and she obtains
with probability $p_k$ the state $\sigma_k$. Then
the entanglement cannot increase on average, i.e:
\bea 
\label{monoLOCC} E(\rho)\geq \sum_k
p_kE(\sigma_k). 
\eea

\item [(b)] Convexity: The entanglement decreases if we discard
some information, i.e. 
\bea 
\label{convex}
E\left[\sum_k p_k \rho_k\right ]\leq \sum_k p_k
E(\rho_k) 
\eea

\end{description}

Now we briefly summarize some useful properties of the
particular measures of entanglement mentioned in the previous
subsection. Let us start with the properties of the concurrence,
$C$, assuming that we have two qubits:

\begin{description}

\item [(i)] $C(\ket{\Psi})=1$ iff
$\mu_k^2=e^{i\delta}|\mu_k|^2$ $(k=1,\ldots,4)$.
This means that a state, written in the magic
basis is maximally entangled iff its coefficients
are real, except for a global phase.

\item [(ii)] $C(\ket\Psi)=0$ iff $\ket{\Psi}$ is a
product state iff $\sum_k \mu_k^2=0$.
\end{description}

These two properties imply that if $\ket{\Phi}$ and
$\ket{\Phi^\perp}$ are real in the magic basis (and therefore
they are maximally entangled) then the state $\ket{\Phi}
\pm i\ket{\Phi^\perp}$ is a product state.

Let us also review some properties of the Entropy of
Entanglement, $E_E$, the Schmidt number, $E_S$, the entanglement
monotones $E_n$, and the R\`{e}nyi entanglement, $E_R$, for
arbitrary states of two $m$--level systems:

\begin{description}

\item [(1)] A maximally entangled state, $\ket{\Psi}$ of
two $m-$level systems has $m$ Schmidt coefficients, which are
all $1/\sqrt{m}$. Thus $E_E(\ket{\Psi})=\log_2(m)$,
$E_S(\ket{\Psi})=m-1$, $E_n(\ket{\Psi})=nE_1=n/m$, ($n=1,\ldots,m-1$)
 and $E_R(\ket{\Psi})=1-1/m$.

\item [(2)] A product state can
be written as
$\ket{\Psi}=\ket{\phi}_A\ket{\psi}_B$, thus it
has only one Schmidt coefficient, which is equal
to $1$. And so
$E_E(\ket{\Psi})=E_S(\ket{\Psi})=E_n(\ket{\Psi})=E_R(\ket{\Psi})=0$
($n=1,\ldots,m-1$).

\end{description}

\section{Unitary operations}

In the next sections we will calculate the maximum attainable
entanglement produced by two--qubit gates. In this Section we
consider an arbitrary unitary operator $U_{AB}$ acting on two
qubits and derive some properties which will simplify the
problem.

In Appendix A we show that for any unitary
operator $U_{AB}$ there exist local unitary
operators, $U_A, U_B, V_A, V_B$, and a unitary
operator $U_d$ such that 
\bea 
U_{AB}=U_A\otimes U_B U_d V_A\otimes V_B, 
\eea 
where 
\bea
\label{Ud} U_d=e^{-i \vec{\sigma_A}^T d
\vec{\sigma_B}} 
\eea 
and $d$ is a diagonal
matrix. Here, $\vec{\sigma_A}^T$ denotes the
transpose of $\vec{\sigma_A}$ expressed in the
computational basis. We will denote the diagonal
elements of $d$ by $\alpha_x,\alpha_y,\alpha_z$.
Note that any measure of entanglement is not
changed by local unitary operators. Thus the
entanglement created by $U_{AB}$ is the same as
the one created by $U_d V_A\otimes V_B$. And so
the maximal amount of entanglement which can be
produced by applying a general unitary, $U_{AB}$
is the same as the one created by $U_d$. This
means that we have to deal with unitaries which
are determined by only $3$ parameters,
($\alpha_x,\alpha_y,\alpha_z$), instead of $15$
parameters, which are used in order to describe a
general unitary operator acting on two qubits.

Furthermore, in Appendix B we show that when studying the
maximum entanglement created by a two-qubit gate we can restrict
ourselves to the case where
\bea
\label{rest}
\pi/4\geq
\alpha_x\geq \alpha_y\geq \alpha_z\geq 0.
\eea
This is due to the fact that the maximal amount of entanglement
created by $U_d$ is symmetric around $\pi/4$ and
$\pi/2$--periodic in $\alpha_x,\alpha_y$ and $\alpha_z$.

It can be easily shown that the operator $U_d$ is diagonal in
the magic basis, and therefore we can write
\bea
U_d=\sum_{k=1}^4 e^{-i\lambda_{k}}\proj{\Phi_k}.
\eea
The phases $\lambda_k$ are
\bea
\lambda_1&=&\alpha_x-\alpha_y+\alpha_z,\\ \nonumber
\lambda_2&=&-\alpha_x+\alpha_y+\alpha_z,\\ \nonumber
\lambda_3&=&-\alpha_x-\alpha_y-\alpha_z,\\ \nonumber
\lambda_4&=&\alpha_x+\alpha_y-\alpha_z. 
\eea

\section{Two qubits}

In this section we consider the following scenario. Alice and
Bob have one qubit each. They want to entangle them by applying
a given unitary operation $U_{AB}$. Their main goal is to find
the best separable (pure\cite{note}) input state that gives as
much entanglement as possible. According to our previous
discussions, we just have to find which states $\ket{\phi}_A,
\ket{\psi}_B$ maximize the concurrence of the output state
$U_{d}\ket{\phi}_A, \ket{\psi}_B$, where $U_d$ is given in
(\ref{Ud}) with restrictions (\ref{rest}). We will call these
states best input states.

Writing the input and output state in the magic
basis with the coefficients $w_k, \mu_k$
respectively, we apply the unitary operator $U_d$
and obtain
\bea \label{eqU1}
\sum_k\mu_k\ket{\Phi_k}=U_d(\ket{\phi}_A\ket{\psi}_B)=
\sum_k w_k e^{-i\lambda_k}\ket{\Phi_k}. 
\eea 
We want to maximize the concurrence of the output
state, $C=|\sum_k \mu_k^2|$, where we have to
make sure that the following conditions are
satisfied:

\begin{description}

\item [(c1)] $\sum_k |\mu_k|^2=1$, which is that the
output state is normalized. Note that, since
$U_d$ is unitary this implies that the input
state is normalized.

\item [(c2)] $\sum_k \mu_k^2 e^{2i\lambda_k}=0$. This condition is due
to the fact that the input state is a product
state, which can be seen as follows. From
Eq.(\ref{eqU1}) we see that $w_k=\mu_k
e^{i\lambda_k}$, and according to Section IIC
(ii) this last one is a product state iff the sum
of the coefficients in the magic basis squared
vanishes, which implies the above equation.

\end{description}

We can determine the maximum of the concurrence
of the output state under the conditions (c1),
(c2) by maximizing $C^2$ and imposing the above
conditions in terms of Lagrange multipliers, i.e.
we maximize 
\bea 
\label{eqC}
f(\mu_1\ldots\mu_4)=\sum_{k,l}\mu_k^2
(\mu_l^\ast)^2-2\eta_1(\sum_k|\mu_k|^2-1) \\ \nonumber 
-\eta_2\sum_k\mu_k^2 e^{2i\lambda_k}-\eta_2^\ast\sum_k(\mu_k^\ast)^2
e^{-2i\lambda_k}, 
\eea 
where $\eta_1$ is real. We
find it convenient to denote $\sum_l
(\mu_l^\ast)^2=Ce^{i\gamma}$, $\eta_2=|\eta_2|
e^{i\epsilon}$, and $\mu_k=|\mu_k|e^{i\xi_k}$. We
obtain 
\bea 
\label{eq1} 
\mu_k\sum_l
(\mu_l^\ast)^2=\eta_1\mu_k^\ast+\eta_2\mu_k
e^{2i\lambda_k} \quad \forall k. 
\eea 
Multiplying
Eq (\ref{eq1}) by $\mu_k$, summing over $k$ and
using (c1) and (c2) we find that $\eta_1=C^2$.
And so we have, assuming that $C\not =0$, 
\bea
\label{eq0} \mu_k(C e^{i\gamma}
 -\eta_2 e^{2i\lambda_k})/C^2=\mu_k^\ast.
\eea 
One of the solutions to this equation is
$\mu_k=0$. To find the others we write Eq.
(\ref{eq0}) as 
\bea 
\label{eq01}
\left|1-\frac{|\eta_2|}{C}e^{i(2\lambda_k-\gamma+\epsilon)}\right|=C.
\eea

Let us distinguish now two cases, namely when $\eta_2$ is zero
or not.

\begin{itemize}

\item $\eta_2=0$: From Eq. (\ref{eq01}) it follows that $C=1$.
Using then Eq. (\ref{eq0}) it is easy to see that
$e^{2i\xi_k}=e^{-i\gamma} \forall k$. Thus all
the coefficients have the same phase (except for the sign) and
therefore the output state is, according to the
discussions of subsection II C, a maximally
entangled state. In order to obtain this state by
applying $U_d$ to a product state the conditions,
(c1) and (c2) still have to be imposed. In
Appendix C we show that those conditions can be
fulfilled iff $\alpha_x+\alpha_y\geq \pi/4$ and
at the same time $\alpha_y+\alpha_z\leq \pi/4$.
There, we also determine the best input state.

\item $\eta_2\not =0$: In this case Eq. (\ref{eq01})
can have at most two solutions for a fixed value
of $|\eta_2|/C$. Thus, in order to fulfill
(\ref{eq1}) $\forall k$ at least two of the
coefficients have to vanish \cite{note2}. Let us
call the other two $\mu_k$ and $\mu_l$. Then, in
order to fulfill conditions (c1) and (c2) we have
to satisfy $|\mu_k|^2+|\mu_l|^2
e^{2i(\lambda_l+\xi_l-(\lambda_k+\xi_k))}=0$ and
the normalization condition. Thus
$|\mu_k|=|\mu_l|=1/\sqrt{2}$ and the difference
between the two phases $\xi_k$ and $\xi_l$ is
$\lambda_l-\lambda_k-\pi/2$. With all that it is
now simple to determine that the largest
reachable concurrence is
\bea 
\label{eq:smax}
C=\mbox{ max}_{k,l} |\sin(\lambda_k-\lambda_l)|.
\eea

Except for global phases the corresponding output
state is $1/\sqrt{2}(\ket{\Phi_k}+i\ket{\Phi_l}
e^{\lambda_k-\lambda_l})$ and the separable input
state, which leads to this maximum is 
\bea
\frac{1}{\sqrt{2}}(\ket{\Phi_k}+i\ket{\Phi_l}).
\eea
Note that the input state $1/\sqrt{2}(\ket{\Phi_k}-i\ket{\Phi_l})$
(the corresponding output state would then be 
$1/\sqrt{2}(\ket{\Phi_k}-i\ket{\Phi_l}
e^{\lambda_k-\lambda_l})$) leads to the same 
amount of entanglement.

\end{itemize}

Note that in case $\alpha_x\le \pi/8$, we obtain that
$C=\sin(\alpha_x+\alpha_y)$, which is directly related to the
entanglement capability of the Hamiltonian of the form
$\sigma_A^Td\sigma_B$ \cite{Du00}. For higher values of
$\alpha_x$ the result may not be directly related to that
quantity.

In summary, in this subsection we have shown that if we apply
$U_d$ to a separable input state and calculate the maximum of
the concurrence of the output state, then we find: If
$\alpha_x+\alpha_y\geq\pi/4$ and $\alpha_y+\alpha_z\leq \pi/4$
then this maximum is equal to $1$. Otherwise it is given by
(\ref{eq:smax}). In addition we determined the best input state
in each of those two cases. Note that, since we were dealing
with two--qubit states we could have taken, according to the
discussions in section II C, any other measure of entanglement
to obtain the same result.

\section{Using Ancillas}

We analyze now whether and how it would be possible to increase
the amount of entanglement of the output state with the help of
auxiliary systems. So, we consider the situation where Alice and
Bob have two qubits each \cite{note3}. Let us denote the
auxiliary qubits by $A'$ and $B'$. We allow input states in
which Alice and Bob's qubits are locally entangled, i.e. of the
form $\ket{\phi}_{AA'}\ket{\psi}_{BB'}$. Then they apply a
non--local unitary transformation, $U_{AB}$ to the qubits $A$
and $B$. The question is, then, for which $\ket{\phi}_{AA'}$ and
$\ket{\psi}_{BB'}$ they are able to reach $\mbox{
max}_{\ket{\phi}_{AA'},\ket{\psi}_{BB'}} E(U_{AB}\otimes
\one_{A'B'} \ket{\phi}_{AA'},\ket{\psi}_{BB'})$, where $E$
denotes some measure of entanglement. In what follows we write
again simply $\ket{\phi}$ ($\ket{\psi}$) instead of
$\ket{\phi}_{AA'}$ ($\ket{\psi}_{BB'}$). On the other hand,
according to Section III, we can restrict ourselves to operators
$U_d$ of the form (\ref{Ud}). For convenience we will call the
input state where $\ket{\phi}$ and $\ket\psi$ are both maximally
entangled, {\it local maximally entangled} and the one where
both are product states, {\it local product state}.

The main difference with the previous section is that now the
best input states depend on the measure of entanglement. To
illustrate this fact we show in the first subsection that for
some measures of entanglement the best input states are the ones
where $\ket{\phi}$ and $\ket\psi$ are entangled. On the other
hand, there are measures of entanglement for which a local
product state is the best input state.

In the second subsection we show that for some special class of
$U_d$ (where $d$ has only one non vanishing element) the
solution to our problem is independent of the measure of
entanglement. In particular, we show how much entanglement can
be created in this case and what is the best input state.
Furthermore, for the class of $U_d$ where all the diagonal
elements of $d$ are the same we will determine the maximum R\`{e}nyi
entanglement as well as the best input state according to this
measure of entanglement.

\subsection{Dependence on the measure of entanglement}

Let us compare the answer to our problem for some
of the measures of entanglement which we recalled
in Section II. According to some numerical
examples we have that:

\begin{itemize}

\item  Schmidt number: Local maximum entangled states are
always better than local product states. This can be easily
understood since in the first case the maximum value which $E_S$
can take is $4$, whereas in the latter one it can be at most
$2$. Thus using this measure of entanglement the ancillas will
in general increase the entanglement of the output state.

\item  R\`{e}nyi entanglement: We have checked that
for this measure the best input states are always either local
product states or local maximally entangled states. In
particular, in the next subsection we will provide analytical
results for some particular cases.

\item Entanglement monotones: We have verified that
there are unitary operators $U_d$ for which local product states
are the best input states, whereas for some other values the
local maximally entangled states lead to the most entangled
output state. But there also exist some $U_d$ for which neither
the product state nor the maximally entangled state are the best
input states.

\end{itemize}

From these examples it becomes clear that it does not make much
sense to ask for the best input state, if one does not specify
according to which measure of entanglement.

\subsection{Examples}

Before we start with the examples let us make some general
statement about the input state. It can always be written in the
Schmidt decomposition as
\bma
\bea
\ket{\phi}_{AA'}&=&c_a
\ket{\phi_0}_A\ket{0}_{A'}+s_a
\ket{\phi_0^\perp}_A\ket{1}_{A'},\\
\ket{\psi}_{BB'}&=&s_b
\ket{\psi_0}_B\ket{0}_{B'}+c_b
\ket{\psi_0^\perp}_B \ket{1}_{B'},
\eea
\ema
where $c_a^2+s_a^2=c_b^2+s_b^2=1$. This is due to the fact that
local unitaries applied to $A'$ and $B'$ do not change the
entanglement (and commute with $U_d$). Let us now treat two
cases in which it is possible to determine the best input state
for some measures of entanglement. The first one should be
viewed as a very simple illustration, whereas the second one is
much more involved.

\subsubsection{Example 1}

Let us consider the following simple unitary operator,
\bea
U_d=e^{-i\alpha S_x}=\cos(\alpha)\one-
i\sin(\alpha)\sigma_x\otimes\sigma_x.
\eea
In this case it is fairly simple to determine the output state.
It has at most $2$ Schmidt coefficients and therefore the state
can be viewed as a state describing two qubits. This implies, as
discussed in II C, that all the measures of entanglement are
equivalent when calculating the optimal states. We take $E_E$.

Let us define $\rho_1$ as the density operator whose
off--diagonal elements are zero, whereas the diagonal elements
are the same as the one of the state $\rho$. Using the fact that
the von Neumann entropy is convex we have that $S(\rho_1)\geq
S(\rho)$. Apart from that, since the problem is symmetric under
exchanging system $(AA')$ with $(BB')$, it is easy to verify
that the states with $\sigma_x\ket{\phi}\propto\ket{\phi^\perp}$
and $\sigma_x\ket{\psi}\propto\ket{\psi^\perp}$ lead to the most
entangled output state. Now, since the states
$\ket{\phi}=\ket{1},\ket{\psi}=\ket{1}$, as well as the states
$\ket{\phi}=\ket{\Phi^+},\ket{\psi}=\ket{\Phi^+}$ fulfill this
condition, both, a local product state and a local maximally
entangled state are the best input states. The maximal entropy
of entanglement which can be obtained is then
\be
\mbox{max}E_E= -\cos(\alpha)^2\log_2[\cos(\alpha)^2]
-\sin(\alpha)^2\log_2[\sin(\alpha)^2].
\ee

\subsubsection{Example 2}

Here we determine the best input state, according to the R\`{e}nyi
entanglement, corresponding to a unitary of the form
\bea
U_d=e^{-i\alpha (S_x+S_y+S_z)}=(\cos(\alpha)^3-
i\sin(\alpha)^3)\one\\ \nonumber
-i\sin(\alpha)\cos(\alpha)e^{i\alpha}(S_x+S_y+S_z),
\eea
where $S_\beta\equiv \sigma_\beta\otimes\sigma_\beta$
($\beta=x,y,z$). Here we have used
$[S_x,S_y]=[S_x,S_z]=[S_y,S_z]=0$. In Appendix D we show that,
according to any measure of entanglement the best input state
can always be written as
\bma
\bea
\ket{\phi}_{AA'}&=&c_a \ket{0}_A\ket{0}_{A'}+s_a
\ket{1}_A\ket{1}_{A'} \\
\ket{\psi}_{BB'}&=&s_b\ket{0}_B\ket{0}_{B'}+c_b
\ket{1}_B \ket{1}_{B'},
\eea
\ema
where $s_a^2+c_a^2=s_b^2+c_b^2=1$.

Let us start by proving that the best input state is either a
local product state or a local maximally entangled state.
Calculating the reduced density operator one finds that
$\rho_A=\rho_1
\oplus \rho_2$ ($\tr(\rho_1 \rho_2)=0$), where $\rho_1,\rho_2$ are $2\times2$-
matrices which depend on $s_a,s_b$ and $\alpha$. It is
straightforward to calculate
$E_R(U_d\ket{\phi}\ket{\psi})=S_R(\rho_A)
=S_R(\rho_1)+S_R(\rho_2)$ and determine its maxima. One finds
that the either the local product states ($s_a^2,s_b^2=0,1$) or
the local maximally entangled states $s_a^2=s_b^2=1/2$ always
lead to a maximum of the R\`{e}nyi entanglement. In the case of a
local product state, it is easy to check that the best one is
$\ket{01}$ (or equivalently $\ket{10}$ \cite{note4}). Let us
denote by $E_R^{\mbox{m.e}}$ ($E_R^{\mbox{p.v}}$) the R\`{e}nyi
entanglement for a local maximally entangled input state
(product state $\ket{01}$) respectively. We obtain
\bma
\bea
E_R^{\mbox{m.e}}(\alpha)&=& \frac{3}{16}[3-2
\cos(4\alpha)-\cos(4\alpha)^2]\\
E_R^{\mbox{p.v}}(\alpha)&=&\frac{1}{2}[1-\cos(4\alpha)^2].
\eea
\ema
Comparing those two expressions we find that
$\alpha_0=\arccos(1/5)/4 \approx 0.109 \pi$. So we have that
$\forall \alpha <\alpha_0$ the local product state is the best
input state and otherwise the local maximally entangled state
leads to the output state with the largest R\`{e}nyi entanglement.
In Fig.\ 1 we illustrate this result.

\begin{figure}[tbp]
\includegraphics[width=8cm]{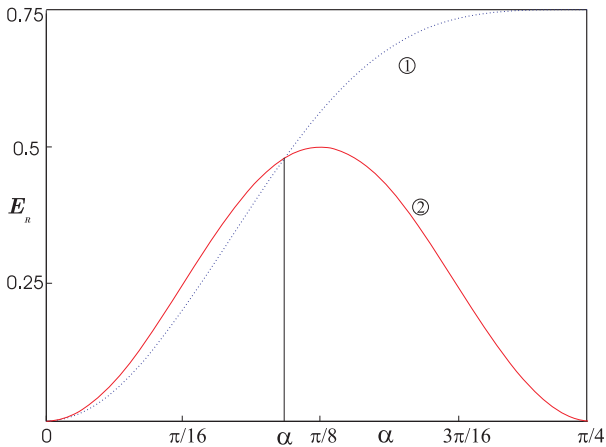}
\caption[]{R\`{e}nyi entanglement for the local
maximally entangled input state ($1$) and for the
product state $\ket{01}$ ($2$)} 
\label{Fig1}
\end{figure}

\section{Conclusions}

We have shown which separable pure two--qubit states have to be
used in order to create as much entanglement as possible by
applying a general two--qubit gate. We have shown which
unitary operators are able to create a maximally entangled state. For
all the other unitary operators we have given the maximal amount
of entanglement which can be created by them
(Eq.(\ref{eq:smax})). Furthermore we have shown that by using
ancillas one has to specify which is the measure of entanglement
to be maximized. We have given two examples of unitary
operations for which it is possible to determine the maximal
amount of some particular measure of entanglement.

\section{Acknowledgments}

We would like to thank Wolfgang D\"ur, Guifre Vidal and,
especially, Maciek Lewenstein for fruitful discussions. This
work has been supported by the Austrian Science Foundation (SFB
``control and measurement of coherent quantum systems''), the
ESF PESC Programm on Quantum Information, TMR network
ERB--FMRX--CT96--0087, the IST Programme EQUIP, and the
Institute for Quantum Information GmbH.

\appendix

\section{Decomposition of unitary operators}

In this appendix we show that for any unitary operator $U_{AB}$
acting on two qubits there exist local unitary operators
$U_A,U_B, V_A,V_B$ and a non--local unitary $U_d=e^{-i\sigma_A^T
d\sigma_B}$ ($d$ diagonal) such that
\bea
\label{eq:decomp} U_{AB}=U_A\otimes U_B U_d
V_A\otimes V_B. 
\eea 
Our prove will be
constructive. Let us call a basis consisting of
maximally entangled orthonormal states a
maximally entangled basis. In what follows the
use of the subscript $k$ implies that the
definition or statement is true for
$k=1,\ldots,4$, if not stated differently.

{\bf Lemma $1$:} For any maximally entangled
basis $\{\ket{\Psi_k}\}$, there exist phases
$\zeta_k$ and local unitaries $U_A, U_B$ such
that 
\bea \label{A2} U_A\otimes U_B e^{i\zeta_k}
\ket{\Psi_k}=\ket{\Phi_k}. 
\eea

{\bf Proof:} According to the discussion in
Subsection IIC (i) we can always write
$\ket{\Psi_k}=e^{\gamma_k}\ket{\bar{\Psi}_k}$,
where $\ket{\bar{\Psi}_k}$ is real in the magic
basis. Let us consider two different states,
$\ket{\bar{\Psi}_k}$ and $\ket{\bar{\Psi}_l}$,
then
$1/\sqrt{2}(\ket{\bar{\Psi}_k}-i\ket{\bar{\Psi}_l})=\ket{e,f}$
and
$1/\sqrt{2}(\ket{\bar{\Psi}_k}+i\ket{\bar{\Psi}_l})=
\ket{\tilde{e},\tilde{f}}$, where
$\ket{e},\ket{\tilde{e}}\in {\cal H}_A$ and
$\ket{f},\ket{\tilde{f}}\in {\cal H}_B$. Note
that $\ket{e,f}$ must be orthogonal to
$\ket{\tilde{e},\tilde{f}}$. This immediately
implies that these vectors must give the Schmidt
decomposition of both $\ket{\bar\Psi_{k,l}}$.
Thus we can write 
\bea
\ket{\bar{\Psi}_1}=\frac{1}{\sqrt{2}}(\ket{e,f}+\ket{e^\perp,f^\perp})\\
\ket{\bar{\Psi}_2}=\frac{-i}{\sqrt{2}}(\ket{e,f}-\ket{e^\perp,f^\perp}).
\eea 
Using the same arguments for
$\ket{\bar\Psi_{3,4}}$ it is easy to determine
that they can be written as 
\bma 
\bea
\ket{\bar{\Psi}_3}&=&\frac{-i}{\sqrt{2}}(e^{i\delta}
\ket{e,f^\perp}+e^{-i\delta}\ket{e^\perp,f})\\
\ket{\bar{\Psi}_4}&=&\pm
\frac{1}{\sqrt{2}}(e^{i\delta}\ket{e,f^\perp}-
e^{-i\delta}\ket{e^\perp,f}) 
\eea 
\ema 
for some $\delta$. In this case, choosing 
\bma 
\bea
U_A&=&\ket{0}\bra{e}+\ket{1}\bra{e^\perp}
e^{i\delta},\\
U_B&=&\ket{0}\bra{f}+\ket{1}\bra{f^\perp}
e^{-i\delta} 
\eea 
\ema 
and the phases $\zeta_k$
appropriately, we have (\ref{A2}). $\Box$.

Note that this first lemma implies that one can go from one
maximally entangled basis to any other using only local
unitaries, if one chooses the global phases appropriately.

{\bf Lemma $2$:} Given a general unitary
operator, then there always exist phases
$\epsilon_k$ and two maximally entangled basis
$\{\ket{\Psi_k}\}$ and $\{\ket{\tilde{\Psi_k}}\}$
such that 
\bea 
U \ket{\Psi_k}=e^{i\epsilon_k}
\ket{\tilde{\Psi_k}}. 
\eea

{\bf Proof:} We give a constructive proof. Let us
denote by $\{\ket{\Psi_k}\}$ the eigenstates of
$U^T U$, where $U^T$ denotes the transpose of $U$
in the magic basis and $e^{2i\epsilon_k}$ are the
corresponding eigenvalues. Note that the
eigenvectors of the symmetric operator $U^T U$
are orthonormal and real, except for global
phases. Thus, since we are working in the magic
basis, they build a maximally entangled basis.
Now we define $\ket{\tilde{\Psi_k}}$ as
\bea
\label{eqU} \ket{\tilde{\Psi_k}}\equiv
e^{-i\epsilon_k}U \ket{\Psi_k}.
\eea

Since the set $\{\ket{\tilde{\Psi_k}}\}$ also
form an orthonormal basis, it remains to prove
that its elements are real. In order to show that
let us consider the eigenvalue equation, $(U^T
U-e^{2i\epsilon_k}\one)\ket{\Psi_k}=0$.
Multiplying it by $U^\ast e^{-i\epsilon_k}$ we
get that $(e^{-i\epsilon_k} U -e^{i\epsilon_k}
U^\ast) \ket{\Psi_k}=0$, which is true iff
$e^{-i\epsilon_k} U \ket{\Psi_k}$ is real $\Box.$

With all that we are now in the position to show that any
unitary operator can be decomposed into local operators and
$U_d$ as in Eq.(\ref{eq:decomp}). So let us now give the
procedure to determine the unitary operators that appear there.

$1$) Calculate the eigensystem of the unitary,
symmetric operator $U^T U$. Let us denote the
eigenvalues by $e^{2i\epsilon_k}$ and the
eigenstates by $\ket{\Psi_k}$. As proven in Lemma
$2$ the set of those states is a maximally
entangled basis.

$2$) Choose $V_A,V_B$ and the phases $\xi_k$, as
explained in Lemma $1$, such that
\bea
\label{eq:V}
 V_A\otimes V_B e^{i\xi_k} \ket{\Psi_k}=
\ket{\Phi_k}.
\eea

$3$) Calculate
\bea
\ket{\tilde{\Psi_k}}=e^{-i\epsilon_k}U\ket{\Psi_k}.
\eea 
Note that according to Lemma $2$ the set of
those states is also a maximally entangled basis.

$4$) Choose the eigenvalues of $U_d$,
$e^{i\lambda_k}$ (note that this is equivalent to
choose the diagonal elements of $d$) and the
unitary operators $U_A, U_B$ such that 
\bea
\label{eq:U} 
U_A^\dagger \otimes U_B^\dagger
e^{i(\lambda_k+\xi_k+\epsilon_k)}
\ket{\tilde{\Psi_k}}=\ket{\Phi_k}, 
\eea 
which, according to Lemma $1$, is always possible. It is
simple to check that with these definitions we
obtain the decomposition (\ref{eq:decomp}).

\section{Periodicity and Symmetry of the maximal amount of
 entanglement}

Let us start out by proving the periodicity of the entanglement
created by $U_d$. We define $d$,($d'$) as a matrix whose
diagonal elements are $\alpha_x,\alpha_y,\alpha_z$
($\alpha_x+\pi/2,\alpha_y,\alpha_z$) respectively. It is simple
to verify that $U_d=-i S_x U_{d'}$. Since $S_x$ is a tensor
product of two local unitary operators the entanglement created
by $U_d$ is the same as the one created by $U_{d'}$. The same
argumentation holds for $\alpha_y$ and $\alpha_z$ and therefore
the amount of entanglement created by $U_d$ is $\pi/2$ periodic
in $\alpha_x,\alpha_y$ and $\alpha_z$.

To prove the symmetry around $ \pi/4$ in
$\alpha_x,\alpha_y, \alpha_z$ of the maximal
amount of entanglement we use the following
definition; $d$, ($d'$) is a matrix whose
diagonal elements are
$\pi/4+\alpha_x,\alpha_y,\alpha_z$ ($\pi/4-
\alpha_x,\alpha_y,\alpha_z$). It is
straightforward to show that $U_d=-i\sigma_x^A
U_{d'}^\ast \sigma_x^B$, where $U_{d'}^\ast$
denotes the complex conjugate of $U_{d'}$ in the
standard basis. And so we have that
$E(U_d\ket{\Psi})= E(U_{d'}^\ast
\sigma_x^B\ket{\Psi})$, where we used that local
unitary operators do not change the entanglement.
Now, we use that for any measure of entanglement,
$E$, $E(\ket{\Psi})=E(\ket{\Psi^\ast})$. This is
obvious, since all the measures are determined by
the Schmidt coefficients and they are real. Thus,
we have that $E(U_d\ket{\Psi})=E(U_{d'}
(\sigma_x^B\ket{\Psi})^\ast)$. It is clear that
the maximal amount of entanglement created by
$U_d$ is the same as the one created by $U_{d'}$.
Again the same argumentation holds for the other
angles, which proves the statement.

\section{Two-qubit gates which create maximally entangled states}

We are going to prove here that there exists a normalized
product state $\ket{\phi}\ket{\psi}$ such that
$\ket{\Phi}=U_d\ket{\phi}\ket{\psi}$ is a maximally entangled
state iff $\alpha_x+\alpha_y\geq \pi/4$ and
$\alpha_y+\alpha_z\leq \pi/4$.

According to our discussions in Subsection II C
and Section IV, this is equivalent to fulfilling
the conditions (c1) and (c2), where
$\mu_k^2=|\mu_k|^2e^{-i\gamma}$. Multiplying
condition (c2) by $e^{-i(\gamma+2\lambda_3)}$, we
obtain
\bea
\label{eq:ii}
|\mu_3|^2+|\mu_1|^2
e^{i\alpha_2}+|\mu_2|^2 e^{i\alpha_3}+|\mu_4|^2
e^{i\alpha_1}=0,
\eea
where we have defined $\alpha_1=4(\alpha_x+\alpha_y)$,
$\alpha_2=4(\alpha_x+\alpha_z)$ and
$\alpha_3=4(\alpha_y+\alpha_z)$. Note that since $\pi/4
\geq\alpha_x\geq\alpha_y\geq\alpha_z\geq 0$, we have that
$2\pi\geq\alpha_1\geq\alpha_2\geq\alpha_3\geq 0$.

Let us distinguish the following two cases now:

\begin{itemize}
\item $\alpha_1 <\pi  $ (or $\alpha_3 > \pi$).
In this case all the imaginary parts appearing in Eq.
(\ref{eq:ii}) are positive (negative) and therefore the sum can
never vanish.

\item $\alpha_1 \geq \pi$ and $\alpha_3\leq \pi$. Here
the imaginary part of $|\mu_4|^2e^{i\alpha_1}$ is negative,
whereas the one of $|\mu_2|^2e^{i\alpha_3}$ is positive and
therefore it is always possible to find a solution to Eq.
(\ref{eq:ii}). In particular we can choose $\mu_1=0$. Then
writing the real and imaginary part of Eq. (\ref{eq:ii}) and the
normalization condition (c1) we simply have to solve:
\bea
\label{eqa}
\sin(\alpha_3)|\mu_2|^2 +\sin(\alpha_1)|\mu_4|^2=0\\
\label{eqb}
|\mu_3|^2+\cos(\alpha_3)|\mu_2|^2+\cos(\alpha_1)|\mu_4|^2=0\\
\label{eqc}
|\mu_2|^2+|\mu_3|^2+|\mu_4|^2=1.
\eea
\end{itemize}

Note that since we have found the solution for
the $\mu_k$'s, it is easy to determine the input
state by using the formula $w_k=\mu_k
e^{i\lambda_k}$.

\section{best input state for example $2$}

Here we prove that the input state which leads to
the most entangled output state can be written as
\bma
\bea
\ket{\phi}&=& c_a\ket{00}+s_a\ket{11}\\
\ket{\psi}&=& s_b\ket{00}+c_b\ket{11},
\eea
\ema
where $s_a^2+c_a^2=s_b^2+c_b^2=1$. We will use that
$[\sigma_{\vec{n}}^A \otimes \sigma_{\vec {n}}^B
,U_d]=0$, where
$\sigma_{\vec{n}}=\vec{\sigma}\cdot \vec{n}$.
This can be easily verified using the commutation
relations of the Pauli operators.

Let us now recall that the input state in system
$AA'$ can be written as $\ket{\phi}=c_a
\ket{\phi_0}_A\ket{0}_{A'}+s_a
\ket{\phi_0^\perp}_A\ket{1}_{A'}$, where
$c_a^2+s_a^2=1$. It is clear that there exists a
vector $\vec{n}$ such that $\sigma_{\vec{n}}
\ket{\phi_0}=\ket{\phi_0}$ and $\sigma_{\vec{n}}
\ket{\phi_0^\perp}=-\ket{\phi_0^\perp}$. Note
that $\ket{\phi}$ is invariant under
$\sigma_{\vec{n}}^A\otimes \sigma_z^{A'}$, i.e.
\bea
\label{eq:phi}
\sigma_{\vec{n}}^A\otimes
\sigma_z^{A'}\ket{\phi}=\ket{\phi}.
\eea

Using the fact that $U_d$ commutes with $\sigma_{\vec{n}}^A
\otimes \sigma_{\vec {n}}^B$ and with local operators acting on
the auxiliary systems together with Eq.(\ref{eq:phi}) we have
that, $\sigma_{\vec{n}}^A\otimes \sigma_z^{A'}\otimes
\sigma_{\vec{n}}^B\otimes V_{B'} U_d \ket{\phi}
\ket{\psi}=U_d \ket{\phi}
\sigma_{\vec{n}}^B\otimes V_{B'}\ket{\psi}$, for
any unitary operator $V_{B'}$.

Let us now introduce a new auxiliary system which we denote by
$C$. Then, using the convexity (\ref{convex}) of any measure of
entanglement, $E$, we have that
\bea
E(U_{AB}\ket{\phi}_{AA'}\ket{\tilde{\psi}}_{BB'C})\geq
E(U_{AB}\ket{\phi}_{AA'}\ket{\psi}_{BB'}),
\eea
where
\be
\ket{\tilde{\psi}}_{BB'C}=1/\sqrt{2}
(\ket{\psi}_{BB'}\ket{0}_C +\sigma_{\vec{n}}^B\otimes V_{B'}
\ket{\psi}_{BB'} \ket{1}_C ).
\ee
Now choosing
$V_{B'}=\sigma_z^{B'}$ and requiring that
$\sigma_{\vec{n}}^B\otimes
V_{B'}\ket{\psi}_{BB'}=\ket{\psi}_{BB'}$, which
implies that $\ket{\psi}= s_b  \ket{\phi_0
0}+c_b\ket{\phi_0^\perp 1}$, we get that
$E(U_{AB}\ket{\phi}_{AA'}\ket{\psi}_{BB'}) \geq
E(U_{AB}\ket{\tilde{\phi}}_{AA'}\ket{\tilde{\psi}}_{BB'})$,
$\forall \ket{\tilde{\phi}},\ket{\tilde{\psi}}$,
where both, $\ket{\phi}$and $\ket{\psi}$ are
invariant under the operation
$\sigma_{\vec{n}}\otimes \sigma_z$. Using the
same argumentation as before, we can apply the
local operator
$\sigma_{\vec{n'}}\otimes\sigma_{\vec{n'}}$,
where $\vec{n'}$ is defined as $\sigma_{\vec{n'}}
\ket{\phi_0}=\ket 0$ and $\sigma_{\vec{n'}}
\ket{\phi_0^\perp}=-\ket 1$. Combining all that
we have that
\bea
E(U_{AB}\ket{\phi}\ket{\psi})\leq
E\{U_{AB}[(c_a\ket{00}+s_a\ket{11})(s_b\ket{00}+c_b\ket{11})]\},
\eea
$\forall$ $\ket{\phi},\ket{\psi}$.



\begin{references}
\bibitem{note}
Note that the maximum entanglement reached for
pure states is always larger than the one reached
by a mixed state. This can be seen as follows:
let us write an arbitrary mixed state, $\rho$ in
its eigenbasis, i.e $\rho=\sum_k p_k
\proj{\Psi_k}$. Using the convexity of any
measure of entanglement, $E$, (\ref{convex}) we
have that $E(\rho)\leq \mbox{max}_{\ket{\Psi_k}}
E(\ket{\Psi_k})$. Thus it suffices to consider
only pure states.
\bibitem{note2}
Note that if two or more $\lambda$'s are equal, this does not
need to be necessary. However, we can discard this case since it
has zero measure in the set of unitary operators, and obtain
this result by imposing continuity.
\bibitem{note3}
Note that since one of Alice' and Bob's subsystems (the ones on
which they apply the unitary operator) are qubits, the state describing
such a subsystem and an ancilla of arbitrary dimension can
always be viewed as a state describing two qubits (it has at
most two Schmidt coefficients).
\bibitem{note4}
Note
that, due to the symmetry of the problem, the entanglement
produced by the input state $\ket{01}$ has to be the same as the
one produced by $\ket{10}$.
\end{references}
\end{document}